\documentclass{article}
\pdfoutput=1
\usepackage{authblk}
\providecommand{\keywords}[1]{\textbf{\textit{Keywords---}} #1}

\usepackage{graphicx}
\setkeys{Gin}{draft=false}

\usepackage{subcaption}

\usepackage{xcolor}
\definecolor{dark-gray}{gray}{0.2}

\usepackage{times}

\usepackage{xspace}
\xspaceaddexceptions{]\}}

\usepackage{environ}

\usepackage{pgfplots}
\usepackage{tikz}
\usetikzlibrary{positioning}
\usetikzlibrary{arrows}

\PassOptionsToPackage{hyphens}{url}
\usepackage[colorlinks,urlcolor=black,linkcolor=dark-gray,citecolor=gray]{hyperref}
\expandafter\def\expandafter\UrlBreaks\expandafter{\UrlBreaks
  \do\a\do\b\do\c\do\d\do\e\do\f\do\g\do\h\do\i\do\j%
  \do\k\do\l\do\m\do\n\do\o\do\p\do\q\do\r\do\s\do\t%
  \do\u\do\v\do\w\do\x\do\y\do\z\do\A\do\B\do\C\do\D%
  \do\E\do\F\do\G\do\H\do\I\do\J\do\K\do\L\do\M\do\N%
  \do\O\do\P\do\Q\do\R\do\S\do\T\do\U\do\V\do\W\do\X%
  \do\Y\do\Z%
}

\usepackage[final]{microtype}

\usepackage{epsfig}

\usepackage{amsmath, amsthm, amssymb, mathtools}
\allowdisplaybreaks

\usepackage{algorithm}
\usepackage{algorithmic}

\usepackage{ifdraft}

\let\oldFootnote\footnote
\newcommand\nextToken\relax
\renewcommand\footnote[1]{%
    \oldFootnote{#1}\futurelet\nextToken\isFootnote}
\newcommand\isFootnote{%
    \ifx\footnote\nextToken\textsuperscript{,}\fi}

\newcommand{\email}[1]{\href{mailto:#1}{\nolinkurl{#1}}}

\newcommand\copyrighttext{%
\footnotesize
To the extent possible under law, Matthieu Vergne has waived all copyright and related or neighboring rights to this technical report. For a more detailed description of this waiving, visit:\\
\url{https://creativecommons.org/publicdomain/zero/1.0/}
}
\newcommand\copyrightnotice{%
\begin{tikzpicture}[remember picture,overlay]
\node[anchor=south,yshift=30pt] at (current page.south) {\fbox{\parbox{\dimexpr\textwidth-\fboxsep-\fboxrule\relax}{\copyrighttext}}};
\end{tikzpicture}%
}

\begin{document}

\title{Mitigation Procedures to Rank Experts\\through Information Retrieval Measures}

\author[1,2]{Matthieu Vergne}

\affil[1]{
  Center for Information and Communication Technology, FBK-ICT\newline
  Via Sommarive, 18 I-38123 Povo, Trento, Italy\newline
  \email{vergne@fbk.eu}
}
\affil[2]{
  Doctoral School in Information and Communication Technology\newline
  Via Sommarive, 5 I-38123 Povo, Trento, Italy\newline
  \email{matthieu.vergne@unitn.it}
}

\date{2016-03-15}

\maketitle

\begin{abstract}
In order to find experts, different approaches build rankings of people, assuming that they are ranked by level of expertise, and use typical Information Retrieval (IR) measures to evaluate their effectiveness.
However, we figured out that expert rankings (i) tend to be partially ordered, (ii) incomplete, and (iii) consequently provide more an order rather than absolute ranks, which is not what usual IR measures exploit.
To improve this state of the art, we propose to revise the formalism used in IR to design proper measures for comparing expert rankings.
In this report, we investigate a first step by providing mitigation procedures for the three issues, and we analyse IR measures with the help of these procedures to identify interesting revisions and remaining limitations.
From this analysis, we see that most of the measures can be exploited for this more generic context because of our mitigation procedures.
Moreover, measures based on precision and recall, usually unable to consider the order of the ranked items, are of first interest if we represent a ranking as a set of ordered pairs.
Cumulative measures, on the other hand, are specifically designed for considering the order but suffer from a higher complexity, motivating the use of precision/recall measures with the right representation.
\end{abstract}

\copyrightnotice
\section{Introduction}
\label{sec:introduction}

When dealing with complex situations, it is sometimes recommended to rely on experts in related domains to obtain the best feedback for achieving our goals.
We work in Requirements Engineering (RE), in which people are asked to identify and maintain the specifications of a system, which often requires to have a broad knowledge about the system and its environment, thus motivating the involvement of domain experts.
To find experts, many systems have been developed with various techniques based on past activities~\cite{mockus_expertise_2002}, question-answer patterns~\cite{zhang_expertise_2007}, language models~\cite{balog_people_2008,serdyukov_modeling_2008}, as well as other and more comprehensive approaches~\cite{mcdonald_expertise_2000,tang_arnetminer:_2008}.
Basically, they aim at building a ranking of people ordered by decreasing level of expertise, allowing to recommend the top people as the most expert.
To evaluate these approaches, mainly inspired from document retrieval techniques, it is common to use Information Retrieval (IR) measures, which compare the rankings produced to some reference rankings.

While we designed our own approach~\cite{vergne_expert_2014}, we noticed that building a reference ranking does not result naturally to a complete and totally ordered ranking~\cite{vergne_gold_2016}.
Yet, these assumptions are common in IR, leading these measures to be hardly applicable to expert rankings unless we enforce them to be complete and totally ordered.
This concern becomes particularly relevant when aiming for Open Source communities, with people freely joining and leaving the community, making infeasible for a human to know exactly who is expert in what in a community of hundreds of members.
This observation leads us to consider that these reference rankings tend to be \emph{incomplete} (all people are not ranked) and \emph{partially ordered} (several people can be at the same rank).
Both these assumptions make the \emph{ranks} of the people a poor indicator to consider, because this rank can change by introducing new people or by introducing additional orders (making it closer to a total order).
At the opposite, the \emph{order} before such modifications is preserved once they are applied, making it a more reliable property to consider.

In this report, we first propose to investigate procedures in Section~\ref{sec:proposal-challenges} to mitigate these new properties, not considered by usual IR measures.
Then, in Section~\ref{sec:art}, we analyse usual IR measures described in the literature by (i) identifying how to make them applicable to our general context by applying the previous procedures and (ii) what are their advantages and limitations.
Finally, in Section~\ref{sec:discussion}, we draw a summary of this analysis highlighting the most interesting measures and the remaining issues for potential future works.


\ifdraft{\clearpage}{}

\section{Problem: Comparing Expert Rankings}
\label{sec:problem}

\paragraph{Context}
We work on a recommender system aiming for recommending the most expert people on some topics, which is achieved by building rankings of experts.
To evaluate our system, we need to compare the rankings produced between each other to assess its stability, or compare them to some references to assess its correctness.
Consequently, we are looking for similarity (or distance) measures to apply for comparing two rankings of experts A and B.
Usually, IR measures are used by relying on the similarity between expert retrieval and document retrieval, but expert rankings have specific properties which need to be considered to design proper comparison measures:
\begin{itemize}
 \item[Incompleteness] Each ranking can be limited in size, especially human-made rankings, leading to have only few people ranked compared to the whole set of available people. Moreover, we can face situations where we should compare a ranking on a subset of people to another ranking on a different (but usually overlapping) subset, which means that we should be able to deal with rankings providing different information.
 \item[Partial order] Some people might be on the same rank, which means that they have the same expertise or, more generally, that we are not able to tell which one is more expert than the other. This can happen even if the human who made the ranking is \emph{more} knowledgeable on the topic than other people, while the latter would provide a more complete order~\cite{vergne_gold_2016}. So we cannot simply assume that a reference ranking must be totally ordered.
 \item[Order consistency] Both these observations lead to a third one, which is that the \emph{rank} is not a reliable property to count on. In case of incompleteness, this rank can vary among rankings on different people, yet the order would be the same (e.g. $a>b>c$ is compatible with $b>c>d$, although $b$ and $c$ have different ranks). Also with partial order, the rank assigned to equal elements implies to consider it like a conflict (e.g. although $a>b>c>d$ is compatible with $a>(b,c)>d$, the equal rank of $b$ and $c$ leads to at least one being in conflict with the former ranking). Moreover, expertise is highly domain-specific~\cite{ericsson_cambridge_2006}, making the identification of the exact level of expertise achieved by a person particularly hard. Most of the time, the people ranked are just ordered based on who appears to be \emph{more or less} expert, independently of their actual level, which again supports that the key property to consider is not the \emph{rank} of a person, but the relative \emph{order} compared to other people.
\end{itemize}

\paragraph{Problem/RQ}
We could design our own measure, but we want first to know if existing measures could be used, which is the purpose of this report.
Thus, we are asking: is there existing similarity/distance measures that can be used to compare expert rankings?
Today, the answer is assumed to be affirmative because expert finding systems are inspired from document retrieval systems~\cite{balog_expertise_2012}, so we use well known IR measures to validate them.
The problem is that, by analysing these measures and how they are used, we can see that they do not fit the three requirements of incompleteness, partial order, and order consistency.


\ifdraft{\clearpage}{}
\section{Mitigation Procedures}
\label{sec:proposal-challenges}

In order to investigate properly existing measures, it is important to consider the specific requirements of our expert rankings: the fact that they represent an \emph{order} only, which can be \emph{partial} and \emph{incomplete}.
One way to do is by finding measures which already consider such properties, which would allow us to build entirely on existing measures, but as we will see in later sections no measure properly deal with all 3 properties.
Another way to increase the chances of finding interesting measures is to consider additional procedures which deal with these properties, so existing measures can be enriched in order to properly satisfy each requirement.
In this section, we present a procedure for each of these 3 properties.

\subsection{Order: Ranking vs. Set of Ordered Pairs}
\label{sec:proposal-challenges-ordinal}

When comparing rankings, it is important to consider the order in which the elements are ranked.
Effort has been made to design measures which look at the order of the ranked items, like CG$_k$ and derived measures (Section~\ref{sec:art-cumulativeGain} and after).
If the measure is already designed for considering orders, then we can use it as-is and deal with expert rankings as an ordered list of people, e.g. $(a, b, c)$.
So the \emph{items} considered by the measure are \emph{people} and the measure directly manage the orders in its own way.

Yet, other measures not considering orders might be of interest, for instance because they are simpler or because the measures considering the orders are based on specific strategies.
If we face such a situation, it is interesting to investigate a way to use it such that we can consider the order.
In our case, we replace a ranking by its equivalent set of ordered pairs, e.g. $a>b>c = \{ a>b, a>c, b>c \}$, so an \emph{item} is an \emph{ordered pair of people} and the measure does not need to consider any order, because they are directly part of the items.
In particular, two rankings based on the same people $r_1 = a>b>c$ and $r_2 = c>b>a$ lead to compare two sets $r_1 = \{ a>b, a>c, b>c \}$ and $r_2 = \{ c>b, c>a, b>a \}$.
Looking at the union of these two sets, like precision (Section~\ref{sec:art-precision}) which does not consider the order, results in an empty set interpreted as a complete difference although they rank the same people, which is what we expect when we look at how the two rankings are reversed.

However, this representation as a set of ordered pairs still has a potential limitation: whether a pair is reversed or absent, the interpretation can be the same.
For instance, if we have :
\begin{itemize}
 \item $r_1 = a>b>c = \{ a>b, a>c, b>c \}$,
 \item $r_2 = c>b>a = \{ c>b, c>a, b>a \}$,
 \item $r_3 = c>b = \{ c>b \}$,
\end{itemize}
then we see that comparing $r_1$ to $r_2$ by looking at the shared pairs ($\emptyset$) gives the same result than comparing $r_1$ to $r_3$, although $r_2$ is closer to $r_1$ because it has the same set of elements.
This problem can be solved by differentiating an absent item from an ``opposite'' item, but none of the measures analysed in this report provide such a feature.

\subsection{Partial Order: Maximal Similarity}
\label{sec:proposal-challenges-partialOrder}

A ranking of experts can be partially ordered simply because one does not have enough information to differentiate the expertise of two people.
In such a case, it is important that the measure used to compare two rankings does not enforce arbitrary assumptions, which is what happen when it requires to have a totally ordered ranking.
For instance, a measure based on the rank of the item might be tricked by a misalignment due to a partial order: $a>b>c>d$ is not conflictual with $a>(b,c)>d$ while it is conflictual with $a>c>b>d$.
Yet, both cases are treated similarly: because $b$ and $c$ have the same rank (2, 2.5 or 3 depending on the method used) then at least one of them is ``wrong'', which is clearly not expected if the reference is the partial one.
Of course, having more than two elements at the same rank makes it even worse, potentially making such a lack of information appearing as more conflictual than a single, but explicit, conflict.

If we face a measure which considers the order of the ranking, we assume that two rankings should be considered as different based on their \emph{explicit} misalignments, so when two items are ordered in a reversed way.
To do so, we consider Algorithm~\ref{algo:computationPartialRankings} which, from two partially ordered rankings, compares all the possible totally ordered rankings and take the maximal similarity.
In the case of a distance, the initial value ($-\infty$) should be reversed ($\infty$) and we should take the minimal value instead of the maximal one.

\begin{algorithm}
\begin{algorithmic}[1]
  \REQUIRE $r_1, r_2$: partially ordered rankings
  \REQUIRE $s$: similarity function for totally ordered rankings
  \ENSURE $sim$: similarity value between $a$ and $b$
  \STATE $R_1 = buildPossibleTotalRankingsFor(r_1)$
  \STATE $R_2 = buildPossibleTotalRankingsFor(r_2)$
  \STATE $sim = -\infty$
  \FORALL{$(r'_1, r'_2) \in R_1 \times R_2$}
    \STATE $sim = max(sim, s(r'_1, r'_2))$
  \ENDFOR
\end{algorithmic}
\caption{Similarity computation for partial rankings.}
\label{algo:computationPartialRankings}
\end{algorithm}

The method $buildPossibleTotalRankingsFor(r)$ provides the set of totally ordered rankings (1 person per rank) which are compatible with the partially ordered ranking $r$.
For instance, for a partial ranking $r = a>(b,c)>d$, the function returns the two possible total rankings $\{ a>b>c>d, a>c>b>d \}$.
If the ranking is already totally ordered, then only one ranking (the same) is returned, and if both rankings are totally ordered, only the similarity measure $s(r_1, r_2)$ is computed, so we go back to the original measure.
This is how we see that this algorithm properly generalizes measures assuming total orders.

\subsection{Incompleteness: Homogenization}
\label{sec:proposal-challenges-partialSet}

Expert rankings are often limited in size: we can restrict them because we are only interested in the top people, or we might face practical limitations, like human-made rankings which cannot rank hundreds of people in a reliable way (e.g. because of fatigue or simply lacks of information).
Of course, human-made rankings can be combined to complete each other, but conflicts can occur~\cite{vergne_gold_2016} and need to be mitigated.
This is usually achieved by assuming that a broad agreement is correct: if a majority of rankings say the same, then it should be true.
Assuming such equivalence is arguable, because to be able to evaluate the expertise of someone, one needs himself to know what to evaluate and how, what is correct, etc. which means to have enough expertise too~\cite{ericsson_cambridge_2006}.
So if, among all the human-made rankings, only few of them are made by actual experts, an agreement-based ranking might loose correct information in favour of social agreements.

Consequently, if the available rankings do not allow to obtain a complete ranking, we need to use a measure able to deal with it.
With this in mind, we might still be interested in measures which assume that the rankings are complete, or at least that rank the same set of people.
In such a case, we need to adapt them to deal with incomplete rankings, which does not mean only having one ranking subset of the other, because each ranking could be incomplete, meaning that each ranks people that the other does not.
For us, if a person is missing for a ranking A, then no constraint is given, so any rank is fine, in particular the rank given by the other ranking B.
In other word, independently of the rank given by B, this person does not intervene in the comparison between the two rankings A and B.
Consequently, we propose to homogenize the two rankings by removing the people which are not in both rankings.


This is consistent with the fact that we consider order rankings (Section~\ref{sec:proposal-challenges-ordinal}): by reducing a ranking, we change the ranks of the lowest people, so measures relying on the rank are impacted, but the order is preserved, so measures considering the order only are not impacted.
In particular, Algorithm~\ref{algo:computationPartialRankings}, which deals with the partial order issue, provides the same result whether or not this procedure is applied.
An important point is that this procedure does not generalizes to more than 2 rankings: it works only because we are making a 1-to-1 comparison.
In the case where we compare 3 rankings A, B, and C, removing a person from A and B because C does not have it means that we ignore any potential misalignment between A and B regarding this person.
Finally, some measures could assume that a missing/extra element should hurt the similarity, in which case no homogenization should be considered to not corrupt the results of the measure.

\ifdraft{\clearpage}{}
\section{Analysis of IR Similarity Measures}
\label{sec:art}

\cite{balog_expertise_2012} presents usual IR measures, established by the TREC community, for evaluating expert finding methods, which are evaluated in exactly the same way as document retrieval systems.
From their point of view, this is a reasonable choice, since ``\emph{the quality of rankings can be estimated independently of what we rank if quality measures for individual items are alike}''.
We confirmed from one of the authors that this sentence essentially means it does not matter whether we rank documents or experts (or other objects), we can use the same measures.
Although we might agree on the feasibility of applying the same measures, we don't see clear evidences that the measures cited are such well-fitted measures for a generic purpose.

In this section, we investigate a broad set of IR measures designed to evaluate how well an IR system performs upon a query $q$ (or a set of queries $Q$).
The measures investigated are based on \cite{balog_expertise_2012} and \cite{manning_introduction_2008}, but also other resources taken from the Web about IR%
\footnote{Introduction to IR: \url{http://nlp.stanford.edu/IR-book/html/htmledition/evaluation-in-information-retrieval-1.html}}%
\footnote{IR on Wikipedia: \url{https://en.wikipedia.org/w/index.php?title=Information_retrieval&oldid=679995615}}%
\footnote{Cumulative Gain and derived measures on Wikipedia: \url{https://en.wikipedia.org/w/index.php?title=Discounted_cumulative_gain&oldid=674619295}}.
For each measure, we provide the definition of the measure and analyse its use for comparing expert rankings, considering also the procedures described in Section~\ref{sec:proposal-challenges} to complete them when necessary.

\NewEnviron{theory}{%
\paragraph{Theory}
\BODY
}

\NewEnviron{application}{%
\paragraph{Application to expert rankings A and B}
\BODY
}

\subsection{Precision (P)}
\label{sec:art-precision}

\begin{theory}
\begin{align*}
P(q) = \frac{|\{\text{relevant items for $q$}\} \cap \{\text{retrieved items for $q$}\}|}{|\{\text{retrieved items for $q$}\}|}
\end{align*}
\end{theory}

\begin{application}
With this measure, we have two types of items: the retrieved ones and the relevant ones.
We can use the ranking A for the retrieved items and B for the relevant ones, such that the intersection of the numerator retrieves the common items between the two, making it a proper similarity measure.
The issue here is that this measure does not consider any order, what can be fixed by using ordered pairs, as described in Section~\ref{sec:proposal-challenges-ordinal}.
However, because there is no difference between absent and reversed pairs, we might be interested in reducing as much as possible the differences caused by absent pairs, especially the ones which \emph{cannot} be shared, which happens for people who are not ranked by both rankings
This can be reduced by restricting the rankings to their shared people by homogenization, as described in Section~\ref{sec:proposal-challenges-partialSet}.

For the remaining pairs, the partial ordering can still lead to absent pairs, which are considered like reversed pairs.
If this behaviour is not wanted, we can reverse the logics by computing the maximal similarity with the procedure described in Section~\ref{sec:proposal-challenges-partialOrder}.
This is a matter of interpretation: we might be interested in using an optimistic measure or a pessimistic one, leading to use the procedure or not.
A better measure might be one which differentiates between explicit agreement (same pair), explicit disagreement (reversed pair), and uncertain agreement (absent pair).

Finally, we can use this measure to compare a ranking to a reference or to compare two rankings in a symmetric way.
In the first case, the reference can be naturally used as relevant items while the other ranking, the one evaluated, can be naturally used as retrieved items.
The difficulty comes with the symmetric comparison: because the numerator considers only one ranking, it implies to have a different normalization factor depending on which ranking is used for what, making it potentially asymmetric.
To make the measure properly symmetric ($P(A,B) = P(B,A)$), we can ensure that both rankings provide the same number of items, which can be achieved by using the three mitigation procedures presented.
If some of them are not used, we can still ensure the symmetry by combining both uses in a symmetric computation, for instance with $\frac{P(A,B) + P(B,A)}{2}$.
\end{application}

\subsection{Recall (R)}
\label{sec:art-recall}

\begin{theory}
\begin{align*}
R(q) = \frac{|\{\text{relevant items for $q$}\} \cap \{\text{retrieved items for $q$}\}|}{|\{\text{relevant items for $q$}\}|}
\end{align*}
\end{theory}

\begin{application}
Recall is in many ways similar to precision, and we can apply the same reasoning to obtain the same conclusions.
The only difference is on the normalization factor, which is the number of relevant items instead of the number of retrieved items, but we can see that by reversing the two we obtain the same formula than for precision ($P(A,B) = R(B,A)$).
Rather than an additional measure, we see that it offers a complementary measure that we already used to make the precision measure symmetric.
Indeed, by computing $\frac{P(A,B) + P(B,A)}{2}$ we actually compute $\frac{P(A,B) + R(A,B)}{2}$ (and also $\frac{R(A,B) + R(B,A)}{2}$).
If precision or recall appear as good similarity measures, one might consider this formula as a good way to combine the strengths of both, especially if we don't use all the mitigation procedures.
\end{application}

\subsection{Fall-out}
\label{sec:art-fallOut}

\begin{theory}
\begin{align*}
\text{fall-out}(q) = \frac{|\{\text{non-relevant items for $q$}\} \cap \{\text{retrieved items for $q$}\}|}{|\{\text{non-relevant items for $q$}\}|}
\end{align*}
\end{theory}

\begin{application}
Like recall offers a complementary perspective to precision by changing the normalization factor, fall-out offers yet another complementary perspective by considering the complement of the relevant items.
In fact, we can use it in the very same way than recall, but to compute a distance measure (as opposed to similarity) because we look at the complement items.
This is possible as long as the items are ordered pairs: if the complement corresponds to the \emph{reversed pairs} (not reversed \emph{and} absent pairs), then we deal with the very same information.
If the items were people only, then the complement would be the non-ranked people (reversing the ranks would have no effect), which is arguably justifiable because the ranked people are usually the \emph{most relevant} people, not the \emph{relevant} ones, so the complement would not be the non-relevant people only.
Anyway, this strategy is not interesting because it does not allow to consider the order of the people, justifying that we use the ordered pairs instead.

Actually, it might be interesting to investigate to which extent recall and fall-out complement each other: although they are strictly opposed for explicit pairs, the absent pairs (present only in the other ranking) are still part of the computation.
In particular, it could be interesting to see if we can find an equivalence between the fall-out and a complemented recall, with or without the maximal similarity procedure described in Section~\ref{sec:proposal-challenges-partialOrder}.
We consider such a deep analysis out of the scope of this report and limit our conclusion to the following: due to their similarity, the fall-out measure shows as much interest as precision and recall as a candidate measure to compare rankings of experts.
\end{application}

\subsection{F-Score}
\label{sec:art-fScore}

\begin{theory}
\begin{align*}
F_\beta(q) = \frac{(1 + \beta^2) (P(q) . R(q))}{\beta^2 P(q) + R(q)}
\end{align*}
$F_{0.5}$, $F_1$ and $F_2$ seems particularly used to, respectively, give priority to precision, have a balanced measure, or give priority to recall.
\end{theory}

\begin{application}
It combines precision and recall and allow to have a symmetric measure with $F_1$, which is interesting when no one is assumed to be a reference.
If another value is used for $\beta$, we might, like mentioned before, compute both $F_\beta(A,B)$ and $F_\beta(B,A)$ and combine them in a balanced way to have a symmetric measure.
Consequently, this measure acts as another solution to combine both precision and recall values, but with an additional complexity which makes it harder to interpret.
\end{application}

\subsection{Precision at \texorpdfstring{$k$}{k} (P@\texorpdfstring{$k$}{k})}
\label{sec:art-kPrecision}

\begin{theory}
\begin{align*}
P@k(q) = \frac{|\{\text{relevant items for $q$}\} \cap \{\text{$k$ first retrieved items for $q$}\}|}{k}
\end{align*}
\end{theory}

\begin{application}
At the opposite of the usual precision computation, P@$k$ aims at evaluating a reduced set of retrieved items, not all of them.
If, like precision, we use a set of ordered pairs, we need to choose the right pairs to consider, which is not straightforward.
Indeed, the rank of a person is described, with this representation, by several pairs, so we need to make a trade-off between the number of pairs to consider for each person and the number of persons represented by these pairs.
Additionally, assuming that a good trade-off would be identified for a given $k$, choosing $k$ might also suffer some arbitrariness.

It seems to us more interesting to use this measure with people as items (rather than pairs) in order to evaluate the correctness of the top of the ranking.
Still, no order is considered by this measure, but we can reduce this issue if the ranking used to represent the \emph{relevant} items is also reduced to the $k$ first items.
In such a case, we actually use the same formula than for precision, but with an additional reduction to $k$ items of both the compared rankings, and evaluating it with an increasing $k$ allows to analyse the full set, as shown for instance with AveP in Section~\ref{sec:art-averagePrecision}.
Nevertheless, the order is only superficially considered and requires to analyse several values.
It also adds to the fact that this measure analyses the rank differences, not only the orders, so we need to homogenize the rankings to avoid the misalignments due to having different people ranked.
\end{application}

\subsection{R-Precision}
\label{sec:art-rPrecision}

\begin{theory}
We compute P@$k$ with $k = |\text{relevant items for $q$}|$, which can be different for each query $q$.
\end{theory}

\begin{application}
This is equivalent to the P@$k$ measure, but applied to a single $k$ value, and with the necessity of knowing the exact number of relevant items, which is different depending on the representation used (ranking or set of pairs).
By using a ranking, the number of relevant items is the size of the ranking providing the relevant items, and with the homogenization, both rankings have the same size and items leading to have always 1, which is useless.
By using a set of ordered pairs, the number of relevant items is the number of pairs, and because of the homogenization, it is finally equivalent to compute the precision as described in Section~\ref{sec:art-precision}.
\end{application}

\subsection{Average Precision (AveP)}
\label{sec:art-averagePrecision}

\begin{theory}
\begin{align*}
AveP(q) = \frac{\sum_{k=1}^{n} \delta_k(q) P@k(q)}{|\text{relevant items for $q$}|}
\end{align*}
with $n$ the number of retrieved items for the query $q$ and $\delta_k(q)$ the relevance of the $k$-th item retrieved (1 if it is relevant, 0 otherwise).
In the case where we rank the whole set of items, so all the relevant items are for sure retrieved, the formula can be rewritten more like a typical weighted average, thus justifying its naming:
\begin{align*}
AveP(q) = \frac{\sum_{k=1}^{n} \delta_k(q) P@k(q)}{\sum_{k=1}^{n} \delta_k(q)}
\end{align*}
\end{theory}

\begin{application}
As we see from the second formula, we compute an average of P@$k$ over the consecutive $k$, which means that the order directly influences the final result.
This is an example of use of P@$k$ as described in Section~\ref{sec:art-kPrecision}, where several values are computed for an analysis able to consider the order.
Indeed, although we consider the same number of P@$k$, by having relevant items sooner in the ranking we consider higher values of P@$k$, leading to a higher value of AveP.
\end{application}

\subsection{Mean Average Precision (MAP)}
\label{sec:art-meanAveragePrecision}

\begin{theory}
\begin{align*}
MAP(Q) = \frac{\sum_{q \in Q} AveP(q)}{|Q|}
\end{align*}
\end{theory}

\begin{application}
With this measure, we change the level of evaluation: while the previous measures allowed us to compare two rankings designed for a specific query $q$, this measure aims at aggregating the values over the whole set of queries $Q$ to evaluate the overall performance of the recommender system.
This strategy can be applied to any query-specific measure described before, and MAP simply applies it to AveP.
In other words, this is not an interesting measure for identifying a comparison function between two rankings, but the general strategy is interesting as an overall measure building on them, so it is not completely out of scope and deserves to be mentioned.
\end{application}

\subsection{Cumulative Gain (CG\texorpdfstring{$_k$}{k})}
\label{sec:art-cumulativeGain}

\begin{theory}
\begin{align*}
CG_k(q)=\sum_{i=1}^k rel_i(q)
\end{align*}
with $k$ the number of top ranks to evaluate and $rel_i(q)$ the relevance of the $i$-th retrieved item for $q$.
Like P@$k$, the aim is to evaluate a reduced amount of items, but the value sums with lower ranks instead of being computed rather independently.
\end{theory}

\begin{application}
From this measure, we consider a different kind of evaluation, where the relevance of an item is provided by an external function $rel$ providing higher values for more relevant items.
With this flexibility, it is possible to consider the orders even with a ranking representation by using a function giving a higher relevance to items which should be closer to the top.
For instance, assuming that one ranking is used as a reference, we can assign a relevance of $1$ to the last item of this ranking and $k$ to the top item, or a more complex distribution to give more weight to some ranks.
Then, CG$_k$ is computed for the other ranking (the retrieved items) based on this reference.
If we need a symmetric value, we can imagine to compute an average of two values: one computing CG$_k$ for the ranking A based on the ranking B, and another for the ranking B based on A.

Although it might seem an interesting measure because the order can be considered through the design of $rel$, the order still has a little impact, because values are summed up on the sole criteria of the presence of the item.
It means that, for the same ranking but re-ordered, the same value is computed, showing that it gives few information individually.
The power of this measure occurs when we compute it for different $k$ values, so we can analyse the evolution of the value in a graph, showing for instance that we gain more relevance when looking at long rankings, so the most relevant items are actually ranked quite low.
Analysing a graph is more complex than analysing a single value, so using this measure requires a greater effort to draw conclusions, which is one of the critics we could made.
Additionally, this measure lacks the required normalization to reliably compare values for one ranking to values for another.

Because this measure relies on the $k$ first items, it means that we should be able to identify what comes ``first'', so using the representation of a set of ordered pairs, which is not ordered, does not seem adapted.
However, if we are computing only the value for $k$ (i.e. the whole set) then all pairs should be considered and no problem of identifying the ``first'' pairs occur.
In this case, we can for example use this representation with a binary function: 1 if the pair is in the right order, 0 otherwise.
Actually, it can be better by assigning a high relevance to pairs in the right order, a low relevance to reversed pairs, and a middle relevance to absent pairs, making the measure able to make the difference between absent and reversed pairs.
This is an advantage of this measure compared to the previous ones, based on precision and recall, which cannot make this difference.
However, we still face the limitation of the normalization, which hurts our ability to compare the values computed for different rankings.
\end{application}

\subsection{Discounted CG (DCG\texorpdfstring{$_k$}{k})}
\label{sec:art-discountedCumulativeGain}

\begin{theory}
\begin{align*}
DCG_k(q) = rel_1(q) + \sum_{i=2}^k \frac{rel_i(q)}{log_2(i)}
\end{align*}
Like CG$_k$, we sum up the relevance values, but this time by applying a logarithmic weight to decrease the global relevance if highly relevant items appear too late in the ranking.
We might take the freedom of slightly changing the logarithmic value in order to rewrite the formula in a simpler way:
\begin{align*}
DCG'_k(q) = \sum_{i=1}^k \frac{rel_i(q)}{log_2(i+1)}
\end{align*}
With such a shape, we see better how it simply enriches the terms of the sum with a weighting factor compared to CG$_k$.
Additionally, it also allows to make the link with another formula used for DCG$_k$:
\begin{align*}
DCG_k(q) = \sum_{i=1}^k \frac{2^{rel_i(q)} - 1}{log_2(i+1)}
\end{align*}
which prioritizes the relevance value upon the weighting factor.
\end{theory}

\begin{application}
At the opposite of CG$_k$, DCG$_k$ directly involves the index of the item in the computation, so we need to map each item to an index.
This addition makes it inapplicable to the representation using sets of ordered pairs because these sets are not ordered (unless we assign the same index to all the pairs, in which case it is equivalent to CG$_k$).
Consequently, we can only use the ranking representation, and the $rel$ function should be designed carefully based on a reference ranking (A or B) before to compute the DCG$_k$ value of the other ranking (resp. B or A).
In this context, DCG$_k$ only revises the weighting strategy, so it does not provide anything more than what CG$_k$ already provides ($rel$ can be directly designed to compute the same values).
In other words, we can summarize DCG$_k$ (and its variants) as a specialization of CG$_k$ applied to rankings with a specific class of $rel$ functions.
In particular, DCG$_k$ does not introduce any normalization allowing to compare values computed for different rankings.
\end{application}

\subsection{Normalized DCG (NDCG\texorpdfstring{$_k$}{k})}
\label{sec:art-normalizedDiscountedCumulativeGain}

\begin{theory}
\begin{align*}
NDCG_k(q) = \frac{DCG_k(q)}{IDCG_k(q)}
\end{align*}
where IDCG$_k$ is the \textit{ideal} DCG$_k$, meaning the value of DCG$_k$ when the items are re-ordered to maximize it (i.e. sorted by decreasing relevance).
\end{theory}

\begin{application}
This measure solves the main issue of the previous ones: the normalization applied allows to compare values computed for different rankings to see which one is better.
As such, what is interesting with this measure is, like for MAP, the generic strategy it involves: divide the value of the ``sub-measure'' (here DCG$_k$) by the best value achievable with this ``sub-measure''.
This integral dependency to the ``sub-measure'' allows to use it even if we use a different one.
In particular, we might be interested in using CG$_k$ on the sets of ordered pairs, leading to have a normalized version following the same strategy:
\begin{align*}
NCG_k(q) = \frac{CG_k(q)}{ICG_k(q)}
\end{align*}
where ICG$_k$ is the \textit{ideal} CG$_k$, meaning the value of CG$_k$ when the pairs retrieved are the same than the reference, or said another way the value of CG$_k$ when we evaluate the ranking A while taking A also as the reference (or evaluating B based on B).
\end{application}

\ifdraft{\clearpage}{}
\section{Discussion}
\label{sec:discussion}

As we saw, many usual IR measures described in Section~\ref{sec:art} are interesting for comparing two rankings of experts, whether we are looking for symmetric or reference-based comparisons.
However, due to the specific properties of our rankings, the usefulness of these measures cannot be highlighted without using the procedures proposed in Section~\ref{sec:proposal-challenges}.
In particular, two main types of measures show up: the measures based on precision and recall, and the cumulative measures.

Precision and recall (sections \ref{sec:art-precision} and \ref{sec:art-recall}), and more extensively with fall-out (Section~\ref{sec:art-fallOut}), show an interesting complementarity worth to exploit.
The fact that they do not consider any order makes them particularly suited for a representation using sets of ordered pairs, described in Section~\ref{sec:proposal-challenges-ordinal}.
Measures derived from them show additional interests: the F-score provides an interesting combination strategy (Section~\ref{sec:art-fScore}), and if P@$k$ has many issues (Section~\ref{sec:art-kPrecision}), particular uses appear to be interesting, like r-precision for ordered pairs (Section~\ref{sec:art-rPrecision}) and AveP for rankings (Section~\ref{sec:art-averagePrecision}).

On the other hand, Cumulative measures like CG$_k$ (Section~\ref{sec:art-cumulativeGain}) use an external relevance function $rel$ which makes them flexible enough to adapt to both representations, rankings as well as sets of ordered pairs.
While one of the main issues of CG$_k$ is its lack of normalization, Section~\ref{sec:art-normalizedDiscountedCumulativeGain} shows how to fix it, and Section~\ref{sec:art-discountedCumulativeGain} shows interesting functions to use, although they are more adapted to rankings than sets of ordered pairs.
However, another main issue of these measures is the need to compute a set of them (for different $k$ values) rather than a single one, at the opposite of precision/recall measures.

Additionally, while all these measures are worth investigating for our purpose, i.e. to compare two rankings of experts, we might argue that it is not enough to evaluate a complete approach, which is the ultimate goal of an evaluation.
Indeed, each ranking corresponds to a specific query $q$, while evaluating an approach implies to analyse it over a full set of queries $Q$.
Nevertheless, usual IR measures already consider such facilities, like for MAP (Section~\ref{sec:art-meanAveragePrecision}) which uses a generic method applicable to any of the described measures (simple average over the set of queries).

Of course, this report remains short in its analysis, its aim being to identify measures worth to investigate rather than making a deep investigation of each of them.
With the procedures proposed in Section~\ref{sec:proposal-challenges}, we were able to highlight potential uses of usual measures for a more general context than what they were designed for.
In particular, by showing how they appear to be equivalent or complementary, and which issues might be faced when using them, this report provides a comprehensive basis for future works.

\ifdraft{\clearpage}{}

\section{Conclusion}
\label{sec:conclusion}

In this report, we investigated usual IR measures to evaluate their applicability to a more generic context than what they were designed for.
In particular, we highlighted the need to consider rankings which are \emph{incomplete} and \emph{partially ordered}, leading to focus on the relative \emph{order} of the ranked items rather than their absolute ranks.
We proposed procedures to adapt IR measures, in order to consider these properties, and shown that most IR measures remain applicable with their help.
In particular, by representing a ranking as a set of ordered pairs, measures based on precision and recall appear as the most interesting measures.
Although cumulative measures (i.e. CG$_k$ and derived) have been designed precisely for advanced analysis, the simplicity of computation and interpretation of the precision/recall measures enforce this regain of interest.

This report, although it considers a comprehensive list of IR measures, remains superficial in its analysis by focusing on specific adaptations and potential uses.
An interesting future work would be a deeper investigation on the consequences of the procedures proposed to adapt these measures, identifying the differences and equivalences, with some highlights on the computation performances.
In particular, it would be interesting to know how usual measures can be rewritten to consider sets of ordered pairs naturally, rather than using additional procedures like homogenization and maximal similarity.
For instance, by replacing the \emph{rank} (or index) of an item in current formalizations by the \emph{number of ordered pairs} which rank this item lower than others, thus relying only on the relative order rather than an absolute rank, it might be possible to generalize these measures and make them able to deal with partial orders and incompleteness more naturally.

\ifdraft{\clearpage}{}

\bibliographystyle{apalike-refs}
\bibliography{citations}
\end{document}